\documentclass[conference]{IEEEtran}
\IEEEoverridecommandlockouts
\usepackage[noadjust]{cite}
\usepackage{amsmath,amssymb,amsfonts}
\usepackage{algorithmic}
\usepackage{graphicx}
\usepackage{textcomp}
\usepackage{xcolor}
\usepackage{stfloats}
\usepackage{algorithm}
\usepackage{algorithmic}
\usepackage{subfigure}
\usepackage{color} 
\newtheorem{Proposition}{Proposition}

\def\BibTeX{{\rm B\kern-.05em{\sc i\kern-.025em b}\kern-.08em
		T\kern-.1667em\lower.7ex\hbox{E}\kern-.125emX}}
\begin{document}
	
	\title{Joint Beamforming and Antenna Movement \\Design for Moveable Antenna Systems \\Based on Statistical CSI}
	\author{\IEEEauthorblockN{Xintai Chen, Biqian Feng, Yongpeng Wu, Derrick Wing Kwan Ng, and Robert Schober}
		\thanks{The work of Y. Wu is supported in part by the National Key R\&D Program of China under Grant 2018YFB1801102, the Fundamental Research Funds for the Central Universities, National Science Foundation (NSFC) under Grant 62122052 and 62071289, 111 project BP0719010, and STCSM 22DZ2229005.}
		\thanks{X. Chen, B. Feng, and Y. Wu are with the Department of Electronic Engineering, Shanghai Jiao Tong University, Minhang 200240, China (e-mail: chenxintai@sjtu.edu.cn; fengbiqian@sjtu.edu.cn; yongpeng.wu@sjtu.edu.cn).}
		\thanks{D. W. K. Ng is with the School of Electrical Engineering and Telecommunications, University of New South Wales, Sydney, NSW, Australia (e-mail: w.k.ng@unsw.edu.au).}
		\thanks{R. Schober is with the Institute for Digital Communications, Friedrich-Alexander University Erlangen-N\"{u}rnberg (FAU), 91058 Erlangen, Germany (e-mail: robert.schober@fau.de).}
		\thanks{Corresponding author: Yongpeng Wu.}
	}
	
	\maketitle
	
	\begin{abstract}
		This paper studies a novel movable antenna (MA)-enhanced multiple-input multiple-output (MIMO) system to leverage the corresponding spatial degrees of freedom (DoFs) for improving the performance of wireless communications. We aim to maximize the achievable rate by jointly optimizing the MA positions and the transmit covariance matrix based on statistical channel state information (CSI). To solve the resulting design problem, we develop a constrained stochastic successive convex approximation (CSSCA) algorithm applicable for the general movement mode. Furthermore, we propose two simplified antenna movement modes, namely the linear movement mode and the planar movement mode, to facilitate efficient antenna movement and reduce the computational complexity of the CSSCA algorithm. Numerical results show that the considered MA-enhanced system can significantly improve the achievable rate compared to conventional MIMO systems employing uniform planar arrays (UPAs) and that the proposed planar movement mode performs closely to the performance upper bound achieved by the general movement mode.
	\end{abstract}
	
	\section{Introduction}
	Advanced multiple-input multiple-output (MIMO) techniques have been proposed to improve the performance of wireless communication systems by leveraging their spatial degrees of freedom (DoFs) \cite{CoMG}-\cite{CLoM}. However, due to the fixed equally spaced positions of the antennas in conventional MIMO systems, it is impossible to fully exploit the spatial variations of the wireless channels across the entire transmit/receive area.
	
	To circumvent this issue, three novel techniques have been proposed: non-uniform antenna arrays (NUAAs) \cite{Nlaa}, \cite{Nadf2}, fluid antennas (FAs) \cite{Meb5}, and movable antennas (MAs) \cite{Mapa}-\cite{MEMC}. These techniques optimize the physical placement of the antennas to enhance the performance of MIMO systems by effectively leveraging the inherent spatial diversity. In the existing literature, NUAAs have been mainly used to tackle the problem of the reduction of the available spatial DoFs resulting from rank-deficient MIMO channels at millimeter wave or terahertz frequencies \cite{Nlaa}, \cite{Nadf2}. However, NUAAs do not always perform better than uniform arrays, since the positions of the antennas cannot be adapted to the dynamic changes of time-varying wireless communication environments. To address this issue, FAs, equipped with a software-controllable fluidic and conductive structure, can conveniently adapt to practical wireless channels by switching the physical position of an antenna flexibly to one of the candidate ports over a fixed-length line space to establishing a favorable channel \cite{Meb5}. However, the positions in FAs can only be switched on a one-dimensional line due to its liquid form, which makes forming antenna arrays difficult. On the other hand, MA-enhanced MIMO systems, where a large number of antennas can be moved continuously in three dimensional space, grant high flexibility for practical array design \cite{Mapa}. As such, a few initial works have studied the performance of MA-enhanced MIMO systems. For instance, in \cite{Mapa}, the maximum channel power gain achieved by a single receive MA was analyzed for deterministic and stochastic channels. Besides, in \cite{Mccf}, an efficient alternating optimization method was developed to maximize the capacity of a point-to-point MIMO system by iteratively optimizing the positions of the transmit and receive MAs as well as the covariance matrix of the transmit signals. In addition, a multiuser communication system employing MAs was investigated in \cite{MEMC}. In particular, by exploiting perfect channel state information (CSI) and zero-forcing and minimum mean-squared error combining, respectively, a gradient descent method was developed to minimize the total transmit power of single-antenna users. However, in MA-enhanced systems it is impractical to acquire the instantaneous CSI for all possible antenna positions due to the delays caused by the antenna movement. None of the existing works on MA-enhanced MIMO systems considered the practical case where only statistical CSI is available.
	
	To fill this gap, this paper investigates the design of a point-to-point MA-enhanced MIMO system based on statistical CSI. The resulting optimization problem is generally intractable as the objective function and the constraints are non-convex and the  expectation over random states is required. The main contributions of this paper can be summarized as follows:
	\begin{itemize}
		\item [1)]
		Firstly, this is the first work that studies the design of MA-enhanced MIMO systems based on statistical CSI. To this end, we adopt the physical-scattering based channel model to characterize the relationship between the channel matrix and the MA positions, which can be used to model any linear channel by appropriately choosing the angular spreads and statistics.
		\item [2)]
		Secondly, we employ the constrained stochastic successive convex approximation (CSSCA) iterative algorithm framework to  effectively obtain a stationary point solution to the formulated design problem for general movement mode. In particular, we obtain a closed-form solution for the transmit covariance matrix in each iteration. Also, we propose two simplified antenna movement modes, namely the linear movement mode and the planar movement mode, for which closed-form solutions for the MA positions in each iteration are obtained.
		\item [3)]
		Finally, we provide numerical results to demonstrate the potential achievable rate gains facilitated by the proposed MA-enhanced systems compared to conventional systems employing fixed-position uniform planar arrays (UPAs). In particular, it is shown that the achievable rate can be significantly improved by optimizing the MA positions and that the proposed planar movement mode achieves a similar performance as the general movement mode, but entails lower computational and hardware complexities.
	\end{itemize}
	
	The remainder of this paper is organized as follows. Section II presents the system model and the problem formulation. Section III provides the proposed CSSCA iterative optimization framework for the general movement mode. Section IV introduces two simplified antenna movement modes. Numerical results and discussions are presented in Section V. Finally, Section VI concludes this paper.
	
	Notations: Symbols for vectors (lower case) and matrices (upper case) are in boldface. $(\cdot)^T$ and $(\cdot)^H$ denote the transpose and conjugate transpose (Hermitian), respectively. The sets of $P \times Q$ dimensional complex and real matrices are denoted by $\mathbb{C}^{P\times Q}$ and $\mathbb{R}^{P\times Q}$, respectively. We adopt $[\boldsymbol{a}]_p$ and $[\boldsymbol{A}]_{p,q}$ to denote the $p$-th entry of vector $\boldsymbol{a}$ and the entry of matrix $\boldsymbol{A}$ in the $p$-th row and $q$-th column, respectively. The trace of matrix $\boldsymbol{A}$ is denoted by $\operatorname{Tr}(\boldsymbol{A})$. $\boldsymbol{A} \succeq \boldsymbol{0}$ indicates that $\boldsymbol{A}$ is a positive semi-definite matrix. $\mathcal{CN}\left(\boldsymbol{0}, \boldsymbol{\Gamma}\right)$ denotes the circularly symmetric complex Gaussian (CSCG) distribution with mean zero and covariance matrix $\boldsymbol{\Gamma}$. $\delta(\cdot)$ denotes the Dirac delta function. $\mathbb{E}\{x\}$ denotes the expected value of $x$. We use $\boldsymbol{I}_K$ to represent a $K$-by-$K$ identity matrix. $\|\boldsymbol{a}\|$ and $\|\boldsymbol{A}\|$ denote the $L_2$-norm of vector $\boldsymbol{a}$ and the Frobenius norm of matrix $\boldsymbol{A}$, respectively. $\left[\boldsymbol{X}\right]^+$ denotes the projection of matrix $\boldsymbol{X}$ onto the positive semidefinite cone. $\left[a, b\right]$ is an interval between $a$ and $b$, while $\left[a\right]$ only contains $a$.
	
	\section{System Model and Problem Formulation}
	\label{subsection: MA-Enhanced MIMO System}
	We consider an MA-enhanced MIMO communication system, where the positions of the antennas can be controlled dynamically by electromechanical drivers such as stepper motors given the channel conditions \cite{Esom, Daso}. We assume that the transmitter (Tx) is equipped with $N$ MAs in a given region $\mathcal{C}_t$ while the receiver (Rx) exploits $M$ MAs in a given region $\mathcal{C}_r$. The coordinates of the transmit and receive MAs are denoted by $\boldsymbol{t}=\left(\boldsymbol{t}_1, \boldsymbol{t}_2, \ldots, \boldsymbol{t}_N\right)^T$ and $\boldsymbol{r}=\left(\boldsymbol{r}_1, \boldsymbol{r}_2, \ldots, \boldsymbol{r}_M\right)^T$, respectively, where $\boldsymbol{t}_n = \left(x_{t,n},y_{t,n}\right) \in \mathcal{C}_t$ and $\boldsymbol{r}_m = \left(x_{r,m},y_{r,m}\right) \in \mathcal{C}_r$ represent the positions of the $n$-th transmit MA and the $m$-th receive MA, respectively. In order to avoid potential coupling between adjacent MAs, a minimum distance $D \geq \lambda/2$ is required between each pair of MAs \cite{AAfL}, i.e., $\left\|\boldsymbol{t}_i-\boldsymbol{t}_j\right\| \geq D$, $\forall i \neq j$, and $\left\|\boldsymbol{r}_i-\boldsymbol{r}_j\right\| \geq D$, $\forall i \neq j$, where $\lambda$ is the wavelength of the carrier. Then, the received signal $\boldsymbol{y}\in\mathbb C^{M\times 1}$ can be written as
	\begin{equation}
		\boldsymbol{y}=\boldsymbol{H}(\boldsymbol{t},\boldsymbol{r},G) \boldsymbol{x}+\boldsymbol{z},
	\end{equation}
	where $\boldsymbol{H}\left(\boldsymbol{t},\boldsymbol{r},G\right) \in \mathbb{C}^{M \times N}$ is the random channel matrix, which depends on the positions of the MAs, $\boldsymbol{t}$ and $\boldsymbol{r}$, and the spatial spreading function $G$ \cite{Dmfc}. $\boldsymbol{x} \in \mathbb{C}^{N \times 1}$ denotes the transmitted vector, which follows a zero-mean Gaussian distribution with covariance matrix $\boldsymbol{Q} \triangleq \mathbb{E}\left\{\boldsymbol{x} \boldsymbol{x}^H\right\}$. $\boldsymbol{z} \sim \mathcal{C N}\left(\boldsymbol{0}, \sigma^2 \mathbf{I}_M\right)$ denotes the additive white Gaussian noise vector at the receiver.	
%	\begin{figure}[t]
%		\centering
%		\includegraphics[width=0.48\textwidth,height=0.2\textwidth]{fig1.pdf}
%		\caption{The MA schematic at the Tx (left-hand side) and Rx (right-hand side).}
%		\label{fig1}
%	\end{figure}
	
	\subsection{Channel Model}
	\label{subsection: Channel Model}
	We assume the far-field channel model holds, where the angles of departure (AoDs)/angles of arrival (AoAs) for different positions in the Tx/Rx regions are identical. 
	$\theta_T\in[0, \pi]$ and $ \phi_T\in[0, \pi]$ denote the elevation and azimuth AoD, respectively, and $\theta_R\in[0, \pi]$ and $ \phi_R\in[0, \pi]$ denote the elevation and azimuth AoA, respectively. Then, the difference in the signal propagation distance between position $\boldsymbol{t}_n$ and the origin of the Tx region in the direction of the AoD $\left(\theta_T, \phi_T \right)$ is given by
	\begin{equation}
		\rho_t(\boldsymbol{t}_n,\theta_T,\phi_T)=x_{t,n} \sin \theta_T \cos \phi_T+y_{t,n} \cos \theta_T.
	\end{equation}
	Accordingly, the resulting phase difference is $\frac{2\pi}{\lambda}\rho_t(\boldsymbol{t}_n,\theta_T,\phi_T)$. Moreover, the array steering vector of the transmit MAs is defined as follows
	\begin{equation}
		\begin{aligned}
			\boldsymbol{a}_T(\boldsymbol{t}, \theta_T , \phi_T) \triangleq & \frac{1}{\sqrt{N}}\left[e^{-j \frac{2 \pi}{\lambda} \rho_t(\boldsymbol{t}_1, \theta_T , \phi_T)}, e^{-j \frac{2 \pi}{\lambda} \rho_t(\boldsymbol{t}_2, \theta_T , \phi_T)},  \right.\\
			&\left. \ldots, e^{-j \frac{2 \pi}{\lambda} \rho_t(\boldsymbol{t}_N, \theta_T , \phi_T)}\right]^T \in \mathbb{C}^{N \times 1}.
		\end{aligned} 
	\end{equation}
	Similarly, the array response vector of the receive MAs is defined as follows
	\begin{equation}
		\begin{aligned}
			\boldsymbol{a}_R(\boldsymbol{r}, \theta_R , \phi_R) \triangleq & \frac{1}{\sqrt{M}} \left[e^{-j \frac{2 \pi}{\lambda} \rho_r(\boldsymbol{r}_1, \theta_R , \phi_R)}, e^{-j \frac{2 \pi}{\lambda} \rho_r(\boldsymbol{t}_2, \theta_R , \phi_R)}, \right. \\ &\left. \ldots, e^{-j \frac{2 \pi}{\lambda} \rho_r(\boldsymbol{t}_M, \theta_R , \phi_R)}\right]^T \in \mathbb{C}^{M \times 1},
		\end{aligned} 
	\end{equation}
	where $\rho_r(\boldsymbol{r}_m,\theta_R,\phi_R)=x_{r,m} \sin \theta_R \cos \phi_R+y_{r,m} \cos \theta_R$ represents the difference in the signal propagation distance between position $\boldsymbol{r}_m$ and the origin of the Rx region in the direction of the AoA $\left(\theta_R, \phi_R \right)$.
	
	Furthermore, the spatial spreading function $G \left( \theta_T , \phi_T , \theta_R , \phi_R \right)$ characterizes the gains of the propagation paths for different spatial directions and is given by
	\begin{equation}
		\begin{aligned}
			&G\left( \theta_T , \phi_T , \theta_R , \phi_R \right) \\
			&= \overline{G}\left( \theta_{T,0}, \phi_{T,0} , \theta_{R,0}, \phi_{R,0} \right) + \widetilde{G}\left( \theta_T, \phi_T, \theta_R, \phi_R\right)\\
			&=c_0 \delta\left(\theta_T-\theta_{T,0}\right) \delta\left(\phi_T-\phi_{T,0}\right)\delta\left(\theta_R-\theta_{R,0}\right)  \\
			&\quad \times \delta\left(\phi_R-\phi_{R,0}\right) + \widetilde{G}\left( \theta_T, \phi_T, \theta_R, \phi_R\right),
		\end{aligned}
	\end{equation}
	where $c_0$ and $\left\{\theta_{T,0}, \phi_{T,0} , \theta_{R,0}, \phi_{R,0}\right\}$ represent the path gain and the angles of the line-of-sight (LOS) component $\overline{G}$, respectively. We assume an uncorrelated Rayleigh scattering environment that is described by a family of zero-mean Gaussian random variables $\left\{\widetilde{G} \left( \theta_T , \phi_T , \theta_R , \phi_R \right) \right\}$ \cite{Dmfc}. Thus, the channel matrix is given as follows
	\begin{equation}
		\label{cm}
		\begin{aligned}	\boldsymbol{H}(\boldsymbol{t},\boldsymbol{r},G)
			\triangleq \overline{\boldsymbol{H}}(\boldsymbol{t},\boldsymbol{r},G) & + \widetilde{\boldsymbol{H}}(\boldsymbol{t},\boldsymbol{r},G),
		\end{aligned}
	\end{equation}
	where $\overline{\boldsymbol{H}}$ and $\widetilde{\boldsymbol{H}}$ are the Rician component and the Rayleigh component, respectively, which are given by
	\begin{equation}
		\label{cm_rice_rayleigh}
		\begin{aligned}	
			\overline{\boldsymbol{H}}(\boldsymbol{t},&\boldsymbol{r},G)
			=c_0 \boldsymbol{a}_R(\boldsymbol{r}, \theta_{R,0}, \phi_{R,0})
			\boldsymbol{a}_T^H(\boldsymbol{t}, \theta_{T,0}, \phi_{T,0}), \\
			\widetilde{\boldsymbol{H}}(\boldsymbol{t},&\boldsymbol{r},G)=\int \int \int \int \widetilde{G}\left( \theta_T, \phi_T, \theta_R, \phi_R\right) \\
			& \times 
			\boldsymbol{a}_R(\boldsymbol{r}, \theta_R , \phi_R)  
			\boldsymbol{a}_T^H(\boldsymbol{t}, \theta_T , \phi_T)  d \theta_T d \phi_T d \theta_R d \phi_R.\\
		\end{aligned}
	\end{equation}
	Without loss of generality, we normalize $\overline{\boldsymbol{H}}$ and $\widetilde{\boldsymbol{H}}$ such that
	\begin{equation}
		\begin{aligned}
			\left\| \overline{\boldsymbol{H}} \right\|^2 = \frac{K}{K+1},
			\left\| \widetilde{\boldsymbol{H}} \right\|^2 = \frac{1}{K+1},
		\end{aligned}
	\end{equation}
	where $K$ is the Rician factor. The channel degrades to a deterministic channel and a Rayleigh fading channel, when $K \rightarrow \infty$ and $K=0$, respectively. 
	
	For a given spatial spreading function $G$, the achievable rate is given by
	\begin{equation}
		\begin{aligned}
			s&\left(\boldsymbol{t}, \boldsymbol{r},\boldsymbol{Q}, G\right) 
			\triangleq \\
			&\log _2 \operatorname{det}\left(\boldsymbol{I}_M +\frac{1}{\sigma^2} \boldsymbol{H}(\boldsymbol{t},\boldsymbol{r},G) \boldsymbol{Q} \boldsymbol{H}^H(\boldsymbol{t},\boldsymbol{r},G)\right).
		\end{aligned}
	\end{equation}
	Hence, the average achievable rate is given by
	\begin{equation}
		\begin{aligned}
			f(\boldsymbol{t}, \boldsymbol{r},\boldsymbol{Q}) & = \mathbb{E}_{G} \left\{s\left(\boldsymbol{t}, \boldsymbol{r},\boldsymbol{Q}, G\right)\right\}.
		\end{aligned}
	\end{equation}
	
	\subsection{Problem Formulation}
	We aim to maximize the achievable rate by jointly optimizing the positions of the MAs and the transmit covariance matrix subject to the position constraints and the power constraint. Accordingly, the proposed optimization problem is formulated as follows
	\begin{subequations}
		\label{op: original}
		\begin{align}
			\max _{\boldsymbol{t}_i\in \mathcal{C}_t, \boldsymbol{r}_j \in \mathcal{C}_r, \boldsymbol{Q} \succeq \boldsymbol{0}} \quad &f\left(\boldsymbol{t}, \boldsymbol{r},\boldsymbol{Q}\right) & \label{op: 1-a}\\
			\text { s.t. } \qquad\quad &\left\|\boldsymbol{t}_i-\boldsymbol{t}_j\right\|^2 \geq D^2,  &\forall i \neq j,\label{op: 1-b}\\
			&\left\|\boldsymbol{r}_i-\boldsymbol{r}_j\right\|^2 \geq D^2,  &\forall i \neq j,\label{op: 1-c}\\
			&\operatorname{Tr}(\boldsymbol{Q}) \leq P, &
		\end{align}
	\end{subequations}
	where $P \geq 0$ is the given maximum transmit power. Solving problem \eqref{op: original} is challenging for the following reasons. Firstly, it is impossible to calculate the exact expected value of objective function $f$ in closed form. Secondly, the position constraints of the transmit and receive MAs are non-convex, which is an obstacle for the development of a computationally efficient solution.
	
	\section{Design of MA-Enhanced MIMO System}
	\label{section: Design of MA-Enhanced MIMO System}
	The optimization problem at hand is generally intractable and obtaining its globally optimal solution is challenging. From the literature \cite{SSCA}, it is known that CSSCA algorithms can acquire a stationary point even if the objective functions and the constraints are non-convex and involve expectations over random states. Thus, we resort to this methodology for solving problem \eqref{op: original} efficiently.

	\subsection{CSSCA Framework}
	In iteration $t$ of the proposed CSSCA algorithm, a new realization of the random spatial spreading function $G^t$ is realized and the objective function in \eqref{op: 1-a} and the constraints in \eqref{op: 1-b}, \eqref{op: 1-c} are replaced respectively by the recursive surrogate functions $\bar f^t\left(\boldsymbol{t}, \boldsymbol{r},\boldsymbol{Q}\right)=\bar f_{t}^t\left(\boldsymbol{t}\right)+\bar f_{r}^t\left(\boldsymbol{r}\right)+\bar f_{Q}^t\left(\boldsymbol{Q}\right)+s\left(\boldsymbol{t}^t, \boldsymbol{r}^t,\boldsymbol{Q}^t, G^t\right)$ and $\bar{g}_{(i,j)}^t\left(\boldsymbol{t}_i,\boldsymbol{t}_j\right)$, $\bar{h}_{(i,j)}^t\left(\boldsymbol{r}_i,\boldsymbol{r}_j\right)$ defined in \eqref{eq: surrogate functions} at the bottom of the next page with initial values $\bar f_{t}^{-1}\left(\boldsymbol{t}\right)=0$, $\bar f_{r}^{-1}\left(\boldsymbol{r}\right)=0$,  and $\bar f_{Q}^{-1}\left(\boldsymbol{Q}\right)=0$, where $a_t^t$, $\boldsymbol{b}_t^t$, $c_t^t$, $a_r^t$, $\boldsymbol{b}_r^t$, $c_r^t$, $a_Q^t$, $\boldsymbol{B}_Q^t$, $c_Q^t$ are the coefficients of the corresponding recursive surrogate functions; $\tau_{\hat s_t}$, $\tau_{\hat s_r}$, $\tau_{\hat s_Q}$, $\tau_{\bar{g}_{(i,j)}}$, $\tau_{\bar{h}_{(i,j)}}$ are negative constants; and sequence $\rho^t \in(0,1]$ is chosen to satisfy $\rho^t > 0$, $\sum_t^{\infty} \rho^t=\infty$, $\sum_t^{\infty} (\rho^t)^2 < \infty$.
	\begin{figure*}[!b]
		\hrulefill
		\begin{equation}
			\label{eq: surrogate functions}
			\begin{aligned}
				\bar f_{t}^t\left(\boldsymbol{t}\right) &= \left(1-\rho^t\right)\bar f_{t}^{t-1}\left(\boldsymbol{t}\right) + \rho^t \left(\nabla_{\boldsymbol{t}}^T s(\boldsymbol{t}^t, \boldsymbol{r}^t,\boldsymbol{Q}^t,G^t) (\boldsymbol{t}-\boldsymbol{t}^t) +\tau_{\hat s_t}\|\boldsymbol{t}-\boldsymbol{t}^t\|^2\right) \\
				&= a_t^t \|\boldsymbol{t}\|^2 +\left(\boldsymbol{b}_t^t \right)^T \boldsymbol{t}+c_t^t, \\
				\bar f_{r}^t\left(\boldsymbol{r}\right) &= \left(1-\rho^t\right)\bar f_{r}^{t-1}\left(\boldsymbol{r}\right) + \rho^t \left(\nabla_{\boldsymbol{r}}^T s(\boldsymbol{t}^t, \boldsymbol{r}^t,\boldsymbol{Q}^t,G^t) (\boldsymbol{r}-\boldsymbol{r}^t) +\tau_{\hat s_r}\|\boldsymbol{r}-\boldsymbol{r}^t\|^2\right) \\
				&= a_r^t \|\boldsymbol{r}\|^2 +\left(\boldsymbol{b}_r^t \right)^T \boldsymbol{r}+c_r^t, \\
				\bar f_{Q}^t\left(\boldsymbol{Q}\right) &= \left(1-\rho^t\right)\bar f_{Q}^{t-1}\left(\boldsymbol{Q}\right) + \rho^t \left(\operatorname{Tr}\left(\nabla_{\boldsymbol{Q}}^H s(\boldsymbol{t}^t, \boldsymbol{r}^t,\boldsymbol{Q}^t,G^t) (\boldsymbol{Q}-\boldsymbol{Q}^t)\right) +\tau_{\hat s_Q}\|\boldsymbol{Q}-\boldsymbol{Q}^t\|^2\right) \\
				&= a_Q^t \|\boldsymbol{Q}\|^2 +\operatorname{Tr} \left(\left(\boldsymbol{B}_Q^t\right) ^H\boldsymbol{Q}\right)+c_Q^t, \\
				\bar{g}_{(i,j)}^t\left(\boldsymbol{t}_i,\boldsymbol{t}_j\right) &=\tau_{\bar{g}_{(i,j)}}\left(\left\|\boldsymbol{t}_i-\boldsymbol{t}_i^t\right\|^2
				+\left\|\boldsymbol{t}_j-\boldsymbol{t}_j^t\right\|^2\right)
				+ 2\left(\boldsymbol{t}_i^t-\boldsymbol{t}_j^t\right)^T\left(\boldsymbol{t}_i-\boldsymbol{t}_j\right)
				-\left\|\boldsymbol{t}_i^t-\boldsymbol{t}_j^t\right\|^2, \\
				\bar{h}_{(i,j)}^t\left(\boldsymbol{r}_i,\boldsymbol{r}_j\right) &=\tau_{\bar{h}_{(i,j)}}\left(\left\|\boldsymbol{r}_i-\boldsymbol{r}_i^t\right\|^2
				+\left\|\boldsymbol{r}_j-\boldsymbol{r}_j^t\right\|^2\right)
				+2\left(\boldsymbol{r}_i^t-\boldsymbol{r}_j^t\right)^T\left(\boldsymbol{r}_i-\boldsymbol{r}_j\right)
				-\left\|\boldsymbol{r}_i^t-\boldsymbol{r}_j^t\right\|^2.
			\end{aligned}
		\end{equation}
	\end{figure*}
	
	According to the CSSCA framework \cite{SSCA}, the optimal solution $\bar{\boldsymbol{t}}^t$, $\bar{\boldsymbol{r}}^t$, $\bar{\boldsymbol{Q}}^t$ in iteration $t$ is obtained by solving problem \eqref{op: surrogate 1} if it is feasible and problem \eqref{op: surrogate 2} otherwise:
	\begin{subequations}
		\label{op: surrogate 1}
		\begin{align}
			\underset{{\boldsymbol{t}_i\in \mathcal{C}_t, \boldsymbol{r}_j \in \mathcal{C}_r, \boldsymbol{Q} \succeq \boldsymbol{0}}}{\operatorname{max}} \quad &\bar f^t\left(\boldsymbol{t}, \boldsymbol{r},\boldsymbol{Q}\right)  \label{op: surrogate f 1}\\
			\text {s.t. } \qquad\quad &\bar{g}_{(i,j)}^t\left(\boldsymbol{t}_i,\boldsymbol{t}_j\right) \geq D^2,  \forall i \neq j, \label{op: surrogate t 1}\\
			&\bar{h}_{(i,j)}^t\left(\boldsymbol{r}_i,\boldsymbol{r}_j\right) \geq D^2,  \forall i \neq j, \label{op: surrogate r 1}\\
			&\operatorname{Tr}(\boldsymbol{Q}) \leq P. \label{op: surrogate Q 1}
		\end{align}
	\end{subequations}
	\begin{subequations}
		\label{op: surrogate 2}
		\begin{align}
			\underset{{\boldsymbol{t}_i\in \mathcal{C}_t, \boldsymbol{r}_j \in \mathcal{C}_r, \boldsymbol{Q} \succeq \boldsymbol{0}}, \alpha}{\operatorname{max}} \quad &\alpha 		\label{op: surrogate alpha 2}\\
			\text {s.t. } \qquad\quad &\bar{g}_{(i,j)}^t\left(\boldsymbol{t}_i,\boldsymbol{t}_j\right) \geq D^2 + \alpha,  \forall i \neq j, \label{op: surrogate t 2}\\
			&\bar{h}_{(i,j)}^t\left(\boldsymbol{r}_i,\boldsymbol{r}_j\right) \geq D^2 + \alpha,  \forall i \neq j, \label{op: surrogate r 2}\\
			&\operatorname{Tr}(\boldsymbol{Q}) \leq P. &
		\end{align}
	\end{subequations}
	
	After obtaining optimal solutions $\bar{\boldsymbol{t}}$, $\bar{\boldsymbol{r}}$, $\bar{\boldsymbol{Q}}$, optimization variables $\boldsymbol{t}$, $\boldsymbol{r}$, $\boldsymbol{Q}$ are updated according to
	\begin{equation}
		\label{eq: update variables}
		\begin{aligned}
			\boldsymbol{t}^{t+1}&=\left(1-\gamma^t\right)\boldsymbol{t}^t+\gamma^t\bar{\boldsymbol{t}}^t,\\
			\boldsymbol{r}^{t+1}&=\left(1-\gamma^t\right)\boldsymbol{r}^t+\gamma^t\bar{\boldsymbol{r}}^t,\\
			\boldsymbol{Q}^{t+1}&=\left(1-\gamma^t\right)\boldsymbol{Q}^t+\gamma^t\bar{\boldsymbol{Q}}^t,\\
		\end{aligned}
	\end{equation}
	where $\left\{\gamma^t \in(0,1]\right\}$ is a decreasing sequence satisfying $\gamma^t \rightarrow 0$, $\sum_t \gamma^t=\infty$, $\sum_t\left(\gamma^t\right)^2<\infty$, $\lim_{t \rightarrow \infty} \gamma^t /\rho^t=0$.
	
	\subsection{Parallel Solution of Problems \eqref{op: surrogate 1} and \eqref{op: surrogate 2}}
	Note that problem \eqref{op: surrogate 1} is equivalent to the following three independent subproblems
	\begin{equation}
		\begin{aligned}
			\bar{\boldsymbol{t}}^t=\underset{\boldsymbol{t}_i \in \mathcal{C}_{t}}{\operatorname{argmax}} & \qquad  \bar f_{t}^t\left(\boldsymbol{t}\right) \\
			\text {  s.t.  } &\qquad\eqref{op: surrogate t 1}, \label{problem_t_S}
		\end{aligned}
	\end{equation}
	\begin{equation}
		\begin{aligned}
			\bar{\boldsymbol{r}}^t=\underset{\boldsymbol{r}_j \in \mathcal{C}_{r}}{\operatorname{argmax}} & \qquad \bar f_{r}^t\left(\boldsymbol{r}\right) \\
			\text {  s.t.  } &\qquad\eqref{op: surrogate r 1}, \label{problem_r_S}
		\end{aligned}
	\end{equation}
	\begin{equation}
		\begin{aligned}
			\bar{\boldsymbol{Q}}^t=\underset{\boldsymbol{Q}\succeq \boldsymbol{0}}{\operatorname{argmax}} & \qquad \bar f_{Q}^t\left(\boldsymbol{Q}\right) \\
			\text {  s.t.  }  &\qquad\eqref{op: surrogate Q 1}. \label{problem_Q_S}
		\end{aligned}
	\end{equation}
	Problems \eqref{problem_t_S} and \eqref{problem_r_S} can be optimally solved by standard convex solvers, such as CVX, while the optimal solution of problem \eqref{problem_Q_S} is given in the following proposition.
	\begin{Proposition}
		\label{prop}
		The optimal solution $\bar{\boldsymbol{Q}}^t$ of \eqref{problem_Q_S} is given by
		\begin{equation}
			\label{optQ}
			\begin{aligned}
				\bar{\boldsymbol{Q}}^t = \boldsymbol{U}_B^t \bar{\boldsymbol{\Lambda}}_Q^t \left(\boldsymbol{U}_B^t\right)^H,
			\end{aligned}
		\end{equation}
		where $\boldsymbol{U}_B^t$ is the eigenmatrix of $\boldsymbol{B}_Q^t$, $\bar{\boldsymbol{\Lambda}}_Q^t = \left[-\frac{1}{2a_Q^t}\left(\boldsymbol{\Lambda}_B^t - u^*\boldsymbol{I}\right)\right]^+$, and $u^* \geq 0$ is an auxiliary variable chosen such that $\operatorname{Tr}(\boldsymbol{\bar\Lambda}_Q^t) = P$. $u^*$ can be found by the bisection method. $\boldsymbol{\Lambda}_B^t$ is a diagonal matrix whose diagonal elements are the eigenvalues of $\boldsymbol{B}_Q^t$.
	\end{Proposition}
	\emph{Proof: }Please refer to the Appendix.
	
	On the other hand, the solution of problem \eqref{op: surrogate 2} is obtained via \eqref{problem_alpha_t_S}-\eqref{problem_alpha_Q_S}:
	\begin{equation}
		\begin{aligned}
			\bar{\boldsymbol{t}}^t=\underset{\boldsymbol{t}_i \in \mathcal{C}_{t},\alpha}{\operatorname{argmax}} &\qquad \alpha_t \\
			\text {  s.t.  }  &\qquad\eqref{op: surrogate t 2},\label{problem_alpha_t_S}
		\end{aligned}
	\end{equation}
	\begin{equation}
		\begin{aligned}
			\bar{\boldsymbol{r}}^t=\underset{\boldsymbol{r}_j \in \mathcal{C}_{r},\alpha}{\operatorname{argmax}} &\qquad \alpha_r \\
			\text {  s.t.  }  &\qquad\eqref{op: surrogate r 2},\label{problem_alpha_r_S}
		\end{aligned}
	\end{equation}
	\begin{equation}
		\begin{aligned}
			\bar{\boldsymbol{Q}}^t=\boldsymbol{Q}^t. \label{problem_alpha_Q_S}
		\end{aligned}
	\end{equation}
	Problems \eqref{problem_alpha_t_S} and \eqref{problem_alpha_r_S} can be optimally solved by standard convex solvers, such as CVX. Then, the optimal value of problem \eqref{op: surrogate 2} is $\alpha=\operatorname{min}(\alpha_t, \alpha_{r})$.
	
	The general movement mode of the MAs considered so far allows for unrestricted movement of each transmit and receive MA within the given region. Although the proposed CSSCA algorithm allows us to efficiently solve problem \eqref{op: original} for the general movement mode, the computational complexity may be high. The high complexity is mainly caused by the nonconvex constraints on the antenna separation in \eqref{op: 1-b} and \eqref{op: 1-c}, which are needed to avoid antenna coupling. Moreover, longer distances for antenna movement may lead to higher delays and increased power consumption. To address these concerns, we propose two simplified antenna movement modes in the following section, aiming to mitigate these challenges.
	
	\section{Antenna Movement Mode Design}
	\label{section: Antenna Move Mode Design}
\begin{figure}[t]
		\centering  %图片全局居中
		\subfigbottomskip=2pt %两行子图之间的行间距
		\subfigcapskip=-5pt %设置子图与子标题之间的距离
		\subfigure[Linear movement mode.\label{fig2: Linear}]{
			\includegraphics[width=0.23\textwidth,height=0.2\textwidth]{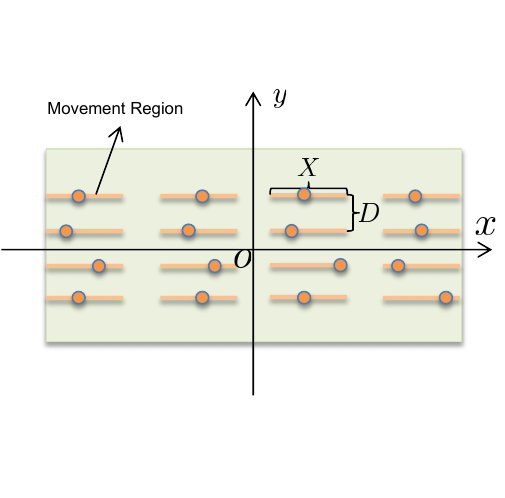}}
		\subfigure[Planar movement mode.\label{fig2: Planar}]{
			\includegraphics[width=0.23\textwidth,height=0.2\textwidth]{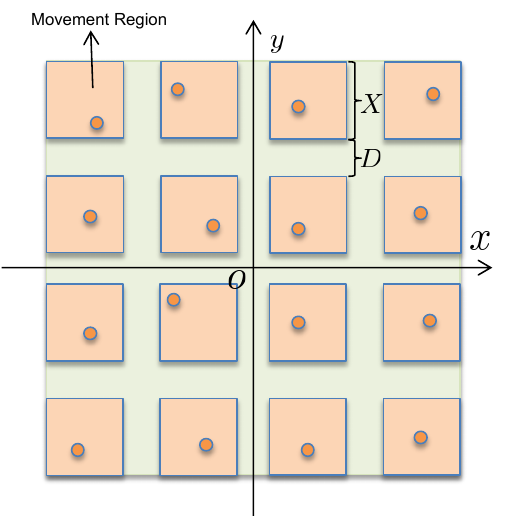}}
		\caption{The proposed antenna movement mode. $D$ is the minimum distance between any two areas. $X$ is the moving range of the MAs in each direction.}
		\label{fig2}
	\end{figure}

	In this section, we propose two simplified movement modes for each MA, namely the linear movement mode and the planar movement mode. As illustrated in Fig.~\ref{fig2}, for these two movement modes, each MA is only allowed to move in given linear or planar area, where the minimum distance between any two areas is set as $D\geq\lambda/2$ to avoid coupling effects. 

	\subsection{Linear  and Planar Movement Mode}
	As shown in Fig.~\ref{fig2: Linear}, for the linear movement mode, the transmit MA $i$ and the receive MA $j$ are allowed to move in the linear areas $\mathcal{C}_{t,i}=[x_{t,i}^{l},x_{t,i}^{u}] \times [y_{t,i}^f]$ and $\mathcal{C}_{r,j}=[x_{r,j}^{l},x_{r,j}^{u}] \times [y_{r,j}^f]$, respectively, where $x_{t,i}^{l}$, $x_{t,i}^{u}$ and $x_{r,j}^{l}$, $x_{r,j}^{u}$ are the lower and the upper bounds of the areas of transmit MA $i$ and receive MA $j$ on the $x$-axis, respectively. $[y_{t,i}^f]$ and $[y_{r,j}^f]$ are fixed positions on the $y$-axis. Hence, the size of movement region for each MA is $X=x_{t,i}^{u}-x_{t,i}^{l}=x_{r,j}^{u}-x_{r,j}^{l}$.
	
%	Thus, function $\mathcal{B}\left\{ \boldsymbol{x} \right\}$ is written as %projects vector $\boldsymbol{x}$ into the corresponding feasible region, i.e.,
%	\begin{equation}
%		[\mathcal{B}\{\boldsymbol{x}\}]_k= 
%		\begin{cases}
%			{[\boldsymbol{x}]_k^l,} & \text { if } k \text{ is odd}, [\boldsymbol{x}]_k<[\boldsymbol{x}]_k^l, \\ 
%			{[\boldsymbol{x}]_k,} & \text { if } k \text{ is odd}, [\boldsymbol{x}]_k^l \leq[\boldsymbol{x}]_k\leq[\boldsymbol{x}]_k^u, \\ 
%			{[\boldsymbol{x}]_k^u,} & \text { if } k \text{ is odd}, [\boldsymbol{x}]_k>[\boldsymbol{x}]_k^u, \\
%			{[\boldsymbol{x}]_k^f,} & \text { if } k \text{ is even},
%		\end{cases}
%	\end{equation}
%	where $[\boldsymbol{x}]_k$ represents the $x$-axis (or $y$-axis) coordinate when $k$ is odd (or even), and thus is projected into $[[\boldsymbol{x}]_k^l,[\boldsymbol{x}]_k^u]$ (or $[[\boldsymbol{x}]_k^f]$). $[\boldsymbol{x}]_k^l$ and $[\boldsymbol{x}]_k^u$ denote the lower and the upper bounds of the feasible region of the $k$-th element of $\boldsymbol{x}$ along the $x$-axis when $k$ is odd, respectively, i.e., $[\boldsymbol{x}]_k^l=x_{t,\frac{k+1}{2}}^{l}$, $[\boldsymbol{x}]_k^u=x_{t,\frac{k+1}{2}}^{u}$ for the transmit MAs and $[\boldsymbol{x}]_k^l=x_{r,\frac{k+1}{2}}^{l}$, $[\boldsymbol{x}]_k^u=x_{r,\frac{k+1}{2}}^{u}$ for the receive MAs. When $k$ is even, $[\boldsymbol{x}]_k^f$ represents fixed $y$-axis coordinate, i.e., $[\boldsymbol{x}]_k^f=y_{t,\frac{k}{2}}^{f}$ for the transmit MAs and $[\boldsymbol{x}]_k^f=y_{r,\frac{k}{2}}^{f}$ for the receive MAs.
	
	As shown in Fig. \ref{fig2: Planar}, for the planar movement mode, transmit MA $i$ and receive MA $j$ are allowed to move in the square areas $\mathcal{C}_{t,i}=[x_{t,i}^{l},x_{t,i}^{u}] \times [y_{t,i}^{l},y_{t,i}^{u}]$ of size $X\times X$ and $\mathcal{C}_{r,j}=[x_{r,j}^{l},x_{r,j}^{u}] \times [y_{r,j}^{l},y_{r,j}^{u}]$ of size $X\times X$, respectively, where $x_{t,i}^{l}$, $x_{t,i}^{u}$, $y_{t,i}^{l}$, $y_{t,i}^{u}$ and $x_{r,j}^{l}$, $x_{r,j}^{u}$, $y_{r,j}^{l}$, $y_{r,j}^{u}$ are the lower and upper bounds on the $x$-axis and $y$-axis of the feasible region of transmit MA $i$ and receive MA $j$, respectively. Since the planar movement mode also allows movement in vertical direction, it can exploit more spatial DoFs but may result in a larger array aperture and a higher cost compared to the linear movement mode.
	
%	Thus, function $\mathcal{B}\left\{ \boldsymbol{x} \right\}$ is written as %projects vector $\boldsymbol{x}$ into the corresponding feasible region, i.e.,
%	\begin{equation}
%		[\mathcal{B}\{\boldsymbol{x}\}]_k= 
%		\begin{cases}
%			{[\boldsymbol{x}]_k^l,} & \text { if } [\boldsymbol{x}]_k<[\boldsymbol{x}]_k^l, \\ 
%			{[\boldsymbol{x}]_k,} & \text { if } [\boldsymbol{x}]_k^l \leq[\boldsymbol{x}]_k\leq[\boldsymbol{x}]_k^u, \\ 
%			{[\boldsymbol{x}]_k^u,} & \text { if } [\boldsymbol{x}]_k>[\boldsymbol{x}]_k^u, \\
%		\end{cases}
%	\end{equation}
%	where $[\boldsymbol{x}]_k$ represents the $x$-axis (or $y$-axis) coordinate when $k$ is odd (or even), and thus it is projected into $[[\boldsymbol{x}]_k^l,[\boldsymbol{x}]_k^u]$. $[\boldsymbol{x}]_k^l$ and $[\boldsymbol{x}]_k^u$ denote the lower and the upper bounds of the feasible region of the $k$-th element of $\boldsymbol{x}$, respectively. 
%	For the transmit MAs, $[\boldsymbol{x}]_k^l=x_{t,\frac{k+1}{2}}^{l}$, $[\boldsymbol{x}]_k^u=x_{t,\frac{k+1}{2}}^{u}$ if $k$ is odd, and $[\boldsymbol{x}]_k^l=y_{t,\frac{k}{2}}^{l}$, $[\boldsymbol{x}]_k^u=y_{t,\frac{k}{2}}^{u}$ otherwise.
%	For the receive MAs, $[\boldsymbol{x}]_k^l=x_{r,\frac{k+1}{2}}^{l}$, $[\boldsymbol{x}]_k^u=x_{r,\frac{k+1}{2}}^{u}$ if $k$ is odd, and $[\boldsymbol{x}]_k^l=y_{r,\frac{k}{2}}^{l}$, $[\boldsymbol{x}]_k^u=y_{r,\frac{k}{2}}^{u}$ otherwise.
	
	\subsection{Simplified Antenna Position Design}
	For the linear or planar movement mode, the optimization problem in \eqref{op: original} reduces to
	\begin{subequations}
		\label{op: original 2}
		\begin{align}
			\max_{\boldsymbol{t}_i\in \mathcal{C}_{t,i}, \boldsymbol{r}_j \in \mathcal{C}_{r,j}, \boldsymbol{Q} \succeq \boldsymbol{0}}   &f\left(\boldsymbol{t}, \boldsymbol{r},\boldsymbol{Q}\right) & \\
			\text { s.t. } \qquad\quad &\operatorname{Tr}(\boldsymbol{Q}) \leq P. &
		\end{align}
	\end{subequations}
	
	Following similar steps as for the general movement mode in Section \ref{section: Design of MA-Enhanced MIMO System}, the optimization of the transmit covariance matrix is also given in the proposition 1 and the optimization of the MA positions in \eqref{problem_t_S}, \eqref{problem_r_S} reduce to
	\begin{align}
		\bar{\boldsymbol{t}}^t&=\underset{\boldsymbol{t}_i \in 	\mathcal{C}_{t,i}}{\operatorname{argmax}} \,\,  \bar f_{S,t}^t\left(\boldsymbol{t}\right),\label{op: special t}\\
		\bar{\boldsymbol{r}}^t&=\underset{\boldsymbol{r}_j \in 	\mathcal{C}_{r,j}}{\operatorname{argmax}} \, \bar f_{S,r}^t\left(\boldsymbol{r}\right).\label{op: special r}
	\end{align}
	The closed-form solutions to the problems in \eqref{op: special t},  \eqref{op: special r}  are given by
	\begin{equation}
		\bar{\boldsymbol{t}}^t=\mathcal{B}\left\{ \frac{\boldsymbol{b}_t^t}{-2a_t^t} \right\},
	\end{equation}
	\begin{equation} \bar{\boldsymbol{r}}^t=\mathcal{B}\left\{ \frac{\boldsymbol{b}_r^t}{-2a_r^t} \right\}, 
	\end{equation}
	respectively. $\frac{\boldsymbol{b}_t^t}{-2a_t^t}$ (or $\frac{\boldsymbol{b}_r^t}{-2a_r^t}$) represents the optimal Tx (or Rx) antenna positioning vector without constraints and function $\mathcal{B}\left\{ \boldsymbol{x} \right\}$ projects vector $\boldsymbol{x}$ into the corresponding feasible region, i.e., $\mathcal{C}_{t,i}$ (or $\mathcal{C}_{r,j}$).
	
	\subsection{Convergence and Complexity Analysis}
	
	\textbf{Convergence analysis:} Every limit point of the sequence $\{\boldsymbol{t}^t,\boldsymbol{r}^t,\boldsymbol{Q}^t\}_{t=1}^\infty$ generated by the CSSCA algorithm is a stationary point for problems \eqref{op: original} and \eqref{op: original 2} almost surely \cite[Thm. 1]{SSCA}.
	
	\textbf{Computational complexity analysis:} The complexity of updating $\boldsymbol{t}$, $\boldsymbol{r}$, and $\boldsymbol{Q}$ mainly depends on the computation of their gradients needed for the approximate problem and the solution of each subproblem. Firstly, the complexities of calculating $\nabla_{\boldsymbol{t}}^T s(\boldsymbol{t}^t, \boldsymbol{r}^t,\boldsymbol{Q}^t,G^t)$, $\nabla_{\boldsymbol{r}}^T s(\boldsymbol{t}^t, \boldsymbol{r}^t,\boldsymbol{Q}^t,G^t)$, and $\nabla_{\boldsymbol{Q}}^T s(\boldsymbol{t}^t, \boldsymbol{r}^t,\boldsymbol{Q}^t,G^t)$ are $\mathcal{O}(M^{3}+NM)$, $\mathcal{O}(M^{3}+NM)$, and $\mathcal{O}(N^{3}+NM)$, respectively. Secondly, for the general movement mode, the complexity of obtaining the solution for each subproblem is $\mathcal{O}(M^{3.5}\log(1/\epsilon)+N^{3.5}\log(1/\epsilon))$ with accuracy $\epsilon$ for the interior-point method. However, the complexity of solving the subproblems for the linear and planar movement modes is $\mathcal{O}(M+N^{3})$, which is much lower than that for the general movement mode. Suppose that the proposed algorithm requires $T$ iterations in total to converge. In this case, the complexity of optimizing $\boldsymbol{t}$, $\boldsymbol{r}$, and $\boldsymbol{Q}$ is $\mathcal{O}(T(M^{3.5}\log(1/\epsilon)+N^{3.5}\log(1/\epsilon)+NM))$ for the general movement mode and $\mathcal{O}(T(M^3+N^3+NM))$ for the linear and planar movement mode.
	
	\section{Numerical Results}
	
%	\begin{figure}
%		\begin{minipage}{1\linewidth}
%			\centering
%			\includegraphics[width=1.0\textwidth,height=0.8\textwidth]{Power_MA.pdf}
%			\caption{Achievable rate versus transmission power.}
%			\label{Simulation: Power}%文中引用该图片代号
%		\end{minipage}
%		\begin{minipage}{1\linewidth}
%			\centering
%			\includegraphics[width=1.0\textwidth,height=0.8\textwidth]{Num_antennas.pdf}
%			\caption{Achievable rate versus the number of transmit/receive antennas.}
%			\label{Simulation: Number of Antennas}%文中引用该图片代号
%		\end{minipage}
%	\end{figure}

	In this section, we numerically evaluate the performance of the proposed MA-enhanced MIMO system for the general movement mode and the two simplified movement modes. To avoid antenna coupling, the minimum distance between adjacent antennas is set as $D=\lambda/2$ in both the MA-enhanced systems and the conventional UPA system. The sizes of the MAs movement areas for the linear and planar movement modes are set as $X=\lambda$ and $X\times X$, respectively. Tx and Rx are both equipped with $M=N=4$ antennas, unless specified otherwise. Besides, the sizes of the entire transmit and receive regions are identical for the general movement mode and the planar movement mode. The spatial spread function $\left\{G \left( \theta_T , \phi_T , \theta_R , \phi_R \right) \right\}$ is simulated as a family of zero-mean Gaussian random variables with non-vanishing values in 4 small angular regions corresponding to scattering clusters with limited angular spreads as they are observed in many realistic environments. The maximum number of iterations for the proposed algorithm is set to 2000.
	\begin{figure}[t]
		\centering
		\includegraphics[width=0.45\textwidth,height=0.35\textwidth]{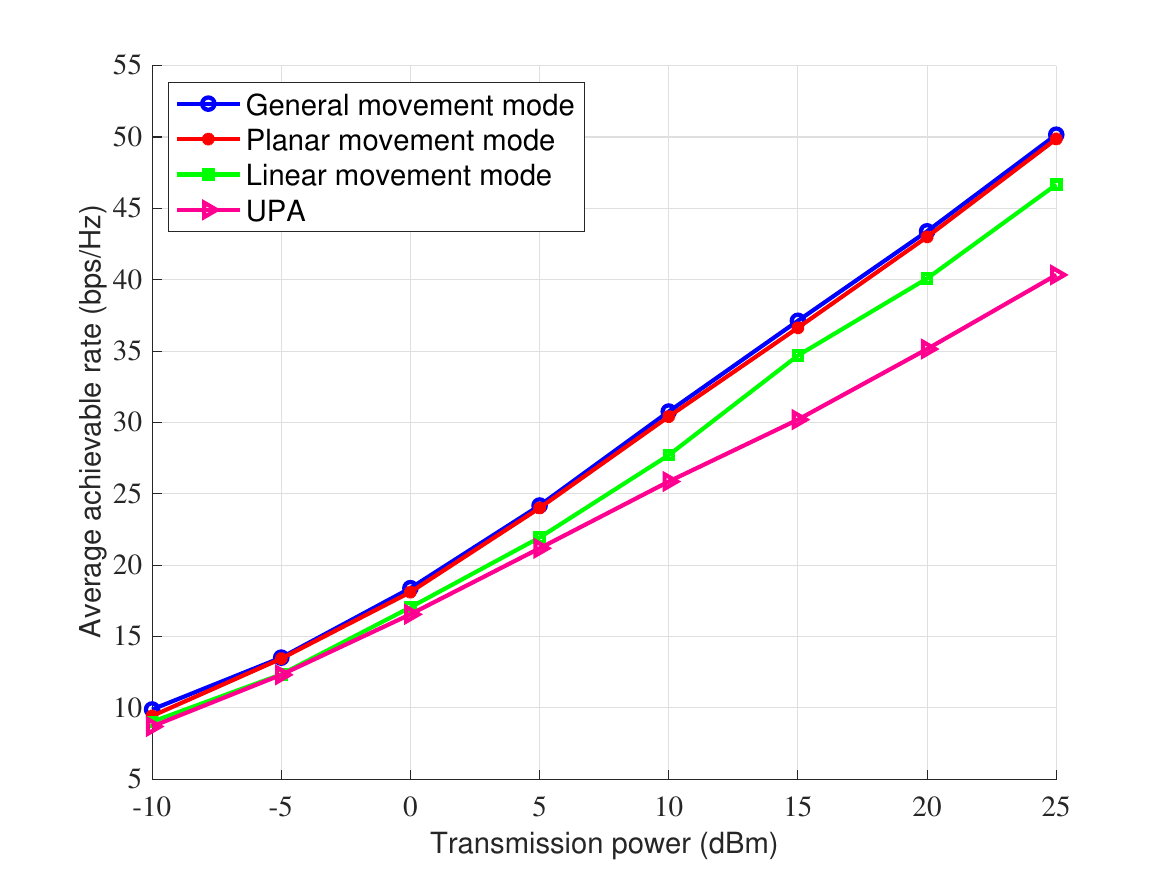}
		\caption{Achievable rate versus transmission power.}
		\label{Simulation: Power}%文中引用该图片代号
	\end{figure}
	\begin{figure}[t]
		\centering
		\includegraphics[width=0.45\textwidth,height=0.35\textwidth]{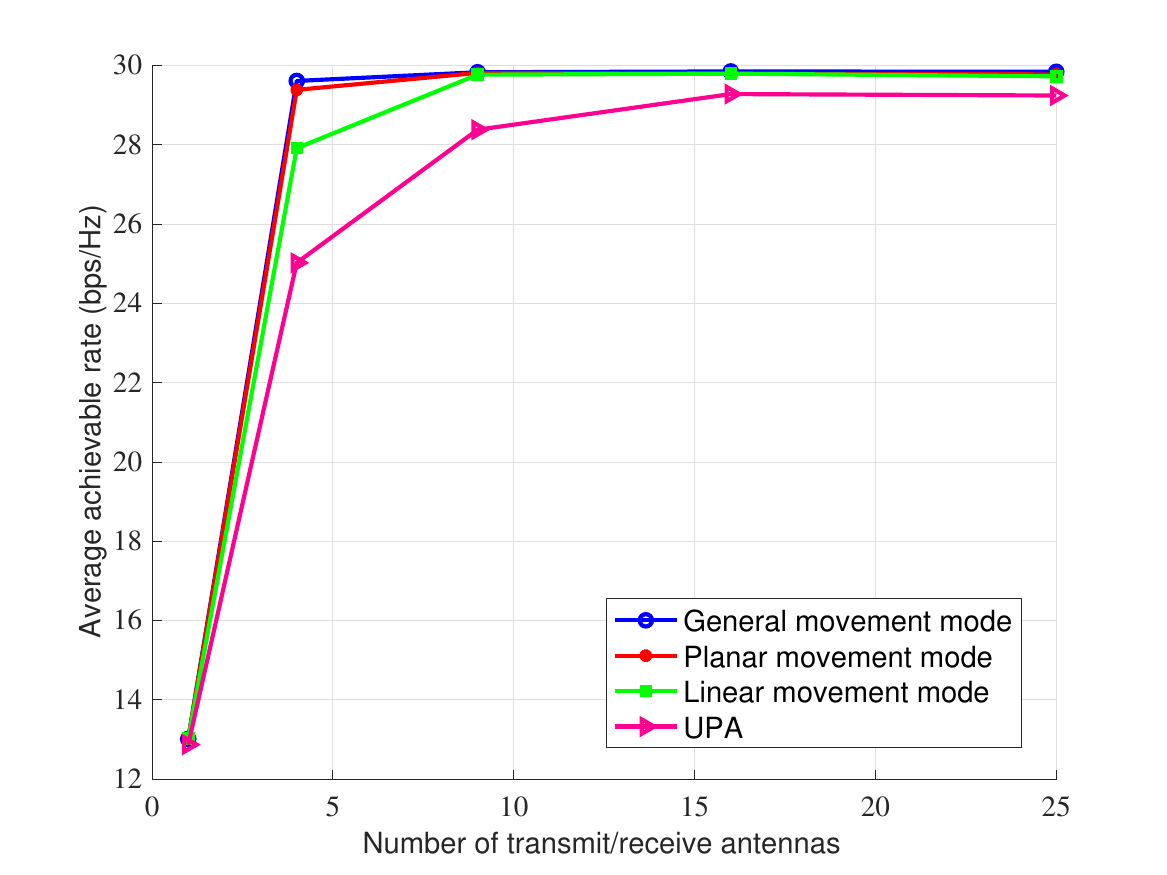}
		\caption{Achievable rate versus the number of transmit/receive antennas.}
		\label{Simulation: Number of Antennas}%文中引用该图片代号
	\end{figure}
	
	\subsection{Impact of Transmit Power}
	Fig.~\ref{Simulation: Power} shows the achievable rate versus the transmit power for different MIMO systems. As expected, for all considered systems, the achievable rate increases monotonically with the transmit power. Moreover, MA-enhanced MIMO outperforms the conventional UPA system, since the MAs can exploit more spatial DoFs. The proposed planar movement mode performs better than the proposed linear movement mode since it can exploit the additional spatial DoFs in the vertical direction. In general, more DoFs result in better performance. In fact, the proposed planar movement mode can closely approach the performance of the general movement mode.	
	
	\subsection{Impact of Number of Antennas}
	Fig.~\ref{Simulation: Number of Antennas} illustrates the impact of the number of transmit/receive antennas on the achievable rate. The performance of the MA-enhanced system equipped with 1 MA at both Tx and Rx does not yield a gain compared to the UPA system as the distribution of the channel coefficient is identical for different positions. On the other hand, the achievable rate for the general and planar movement modes saturates if the number of transmit/receive MAs exceeds 4 due to the limited spatial DoFs introduced by the 4 scattering clusters. MAs can exploit more spatial DoFs than UPAs by adjusting their positions to the radio environment. This highlights the huge potential of MA-enhanced MIMO to improve the future communications systems.
	
	\section{Conclusion}
	In this paper, we investigated a point-to-point MA-enhanced MIMO system with statistical CSI, where Tx and Rx are equipped with multiple MAs. %Both statistical CSI and \ CSI availability were considered in the proposed MA-enhanced MIMO systems. Statistical CSI availability were considered in the proposed MA-enhanced MIMO systems. 
	We applied the CSSCA algorithm framework for maximization of the achievable rate via the joint optimization of the transmit and receive MA positions and the transmit covariance matrix. Moreover, two simplified antenna movement modes, namely the linear movement mode and the planar movement mode, were proposed to facilitate efficient positioning of the MAs with low computational and hardware complexity. Simulation results confirmed the huge potential of MA-enhanced MIMO systems to improve the performance of future communications systems.
	
	\appendix
	Let the optimal $\boldsymbol{Q}$ be expressed as  $\boldsymbol{Q} = \boldsymbol{U}_B^t \boldsymbol{X} \left(\boldsymbol{U}_B^t\right)^H$, where $\boldsymbol{X}$ satisfies $\boldsymbol{X} \succeq \boldsymbol{0}$ and $\operatorname{Tr} \left(\boldsymbol{X}\right) \leq P$, then the smooth objective function is equivalent to
	\begin{equation}
		\begin{aligned}
			\bar f_{Q}^t\left(\boldsymbol{Q}\right)=& a_Q^t \|\boldsymbol{Q}\|^2 +\operatorname{Tr} \left(\left(\boldsymbol{B}_Q^t\right) ^H\boldsymbol{Q}\right)+c_Q^t \\
			=& a_Q^t \|\boldsymbol{X}\|^2 
			+ \operatorname{Tr} \left( \boldsymbol{\Lambda}_B^t \boldsymbol{X} \right)+c_Q^t \\
			\triangleq & \bar f_{\Lambda}^t\left(\boldsymbol{X}\right). \\
		\end{aligned} \label{lambdaQ}
	\end{equation}
	Note that $\bar f_{\Lambda}^t\left(\boldsymbol{\Lambda}_Q\right) \geq \bar f_{\Lambda}^t\left(\boldsymbol{X}\right)$, where $\boldsymbol{\Lambda}_Q$ is a diagonal matrix with diagonal elements $\left[\boldsymbol{\Lambda}_Q\right]_{i,i}=\left[\boldsymbol{X}\right]_{i,i}\geq0$. Thus, problem \eqref{problem_Q_S} is equivalent to
	\begin{equation}
		\label{problem_LambdaQ}
		\begin{aligned}
			\bar{\boldsymbol{\Lambda}}_Q^t=\underset{\boldsymbol{\Lambda}_Q\succeq \boldsymbol{0}}{\operatorname{argmax}} \quad &
			\bar f_{\Lambda}^t\left(\boldsymbol{\boldsymbol{\Lambda}}_Q\right) \\
			\text {s.t.} \qquad & \operatorname{Tr}(\boldsymbol{\Lambda}_Q) \leq P. 
		\end{aligned} 
	\end{equation}
	%And the Karush-Kuhn-Tucker (KKT) conditions of problem \eqref{problem_Q_S} leads to the desired conclusion.
	The partial Lagrangian function of \eqref{problem_LambdaQ} is
	\begin{equation}
		\begin{aligned}
			L\left(\boldsymbol{\Lambda}_Q\right) &= a_Q^t \|\boldsymbol{\Lambda}_Q\|^2 
			+ \operatorname{Tr} \left( \boldsymbol{\Lambda}_B^t \boldsymbol{\Lambda}_Q \right) \\
			&+c_Q^t - u\left(\operatorname{Tr}(\boldsymbol{\Lambda}_Q) - P\right),
		\end{aligned}
	\end{equation}
	where $u \geq 0$ is the Lagrange multiplier. The Karush-Kuhn-Tucker (KKT) conditions reveal that
	\begin{equation}
		\begin{aligned}
			&2a_Q^t \boldsymbol{\Lambda}_Q + \boldsymbol{\Lambda}_B^t - u\boldsymbol{I}=0,\\
			&\boldsymbol{\Lambda}_Q \succeq \boldsymbol{0}, \operatorname{Tr}(\boldsymbol{\Lambda}_Q) \leq P,\\
			&u\left(\operatorname{Tr}(\boldsymbol{\Lambda}_Q) - P\right) = 0, u\geq 0.
		\end{aligned}
	\end{equation}
	This leads to the desired conclusion.

\end{document}